\documentclass[11pt,a4paper]{article}
\usepackage{pos}

\usepackage[utf8]{inputenc}
\usepackage{amsmath}
\usepackage{mathrsfs}
\usepackage{graphicx}
\usepackage{bbold}
\usepackage{comment}
\usepackage{here}
\usepackage{braket}
\usepackage{bbm}
\usepackage{comment}
\usepackage{tikz}
\usepackage{multirow}
\numberwithin{equation}{section}

\newcommand{\be}{\begin{equation}}
\newcommand{\ee}{\end{equation}}

\newcommand{\del}{\partial}

\title{Numerical evidence for a CP broken deconfined phase at $\theta =\pi$ in 4D SU(2) Yang-Mills theory through simulations at imaginary $\theta$}
\ShortTitle{Numerical evidence for a CP broken deconfined phase at $\theta =\pi$ in 4D SU(2)}

\author*[a,b]{Mitsuaki Hirasawa}
\author[c,d,e]{Masazumi Honda}
\author[c,d]{Akira Matsumoto}
\author[f,g]{Jun Nishimura}
\author[h]{Atis Yosprakob}

\affiliation[a]{
Department of Physics, University of Milano-Bicocca,\\
Piazza della Scienza 3, I-20126 Milano, Italy}

\affiliation[b]{
Istituto Nazionale di Fisica Nucleare (INFN), Sezione di Milano-Bicocca,\\
Piazza della Scienza 3, I-20126 Milano, Italy}

\affiliation[c]{
Yukawa Institute for Theoretical Physics, Kyoto University,\\
Kitashirakawa Oiwakecho, Sakyo-ku, Kyoto 606-8502 Japan}

\affiliation[d]{
Interdisciplinary Theoretical and Mathematical Sciences Program (iTHEMS), RIKEN,\\
2-1 Hirosawa, Wako, Saitama 351-0198 Japan}

\affiliation[e]{
Graduate School of Science and Engineering, Saitama University,\\
255 Shimo-Okubo, Sakura-ku, Saitama 338-8570, Japan}

\affiliation[f]{
KEK Theory Center, High Energy Accelerator Research Organization (KEK),\\
1-1 Oho, Tsukuba, Ibaraki 305-0801, Japan}

\affiliation[g]{
Graduate Institute for Advanced Studies, SOKENDAI,\\
1-1 Oho, Tsukuba, Ibaraki 305-0801 Japan}

\affiliation[h]{
Department of Physics, Niigata University,\\
8050 Ikarashi, 2-no-cho, Nishi-ku, Niigata 950-2181 Japan}

\emailAdd{mitsuaki.hirasawa(at)mib.infn.it}
\emailAdd{masazumi.honda(at)yukawa.kyoto-u.ac.jp}
\emailAdd{akira.matsumoto(at)yukawa.kyoto-u.ac.jp}
\emailAdd{jnishi(at)post.kek.jp}
\emailAdd{ayosp(at)phys.sc.niigata-u.ac.jp}
\abstract{
We investigate the possibility of the spontaneous breaking of CP symmetry in 4D SU(2) Yang-Mills at $\theta=\pi$,
which has recently attracted much attention in the context of the higher-form symmetry and the 't Hooft anomaly matching condition.
Here we provide a numerical evidence that the CP symmetry is indeed spontaneously broken at low temperature and it gets restored above the deconfining temperature at $\theta=\pi$,
which is consistent with the anomaly matching condition and yet differs from the situation predicted in the large-$N$ limit.
We avoid the severe sign problem by performing simulations at imaginary $\theta$.
We obtain the critical temperature of the CP restoration and that of deconfinement at $\theta=\pi$ by analytic continuation, which leads to the above conclusion.
}

\FullConference{The 41st International Symposium on Lattice Field Theory (LATTICE2024)\\
 28 July - 3 August 2024\\
Liverpool, UK\\}

\begin{document}
\begin{flushright}
KEK-TH-2687, YITP-25-14, RIKEN-iTHEMS-Report-25, STUPP-24-276
\end{flushright}

\maketitle

\section{Introduction}

4D SU($N$) Yang-Mills theory allows us to add the topological $\theta$ term to the action without breaking Lorentz symmetry and the SU($N$) gauge symmetry.
In general, CP symmetry is explicitly broken at non-zero $\theta$.
However, $\theta = \pi$ is a special point in which the theory has CP symmetry due to the $2\pi$ periodicity in $\theta$.
In particular, the CP symmetry at $\theta=\pi$ is considered to be spontaneously broken at low temperature, while it is known to be restored at sufficiently high temperature \cite{Gross:1980br,Weiss:1980rj}.
The recent study of the model from the perspective of the higher-form symmetry has given a prediction that either the CP or the $\mathbb{Z}_N$ center symmetry should be spontaneously broken unless the theory becomes gapless due to the 't Hooft anomaly matching condition. Therefore, there is a relation between two phase transitions:
\begin{equation}
  T_{\mathrm{CP}} \geq T_{\mathrm{dec}}(\pi) \ ,
  \label{inequality-CP-dec}
\end{equation}
where $T_{\mathrm{CP}}$ represents the temperature at which the CP symmetry at $\theta=\pi$ gets restored, and $T_{\mathrm{dec}}(\theta)$ represents the critical temperature of the deconfining transition, which depends on $\theta$ in general.

In the large-$N$ limit, it is known that these phase transitions occur at the same temperature $T_{\mathrm{CP}} = T_{\mathrm{dec}}(\pi)$ \cite{Witten:1980sp,Witten:1998uka,Bigazzi:2015bna}.
On the other hand, it is predicted that $T_{\mathrm{CP}} > T_{\mathrm{dec}}(\pi)$ at $N=2$ from the analysis of the SU($N$)
supersymmetric Yang-Mills theory (SYM) 
deformed by the gaugino mass and compactified on $S^1$ with periodic boundary conditions regarding the radius of $S^1$ as an analog of the inverse temperature \cite{Chen:2020syd}.
Note that the gaugino mass has to be small enough to make the analysis based on supersymmetry reliable although the theory becomes equivalent to pure Yang-Mills theory in the infinite mass limit.
Note also that the radius of $S^1$ with periodic boundary conditions cannot be regarded as the inverse temperature, which actually requires anti-periodic boundary conditions for the gaugino field.
These subtleties give us a strong motivation to investigate the phase structure 
of 4D SU($2$) pure Yang-Mills theory by numerical methods.


Since the $\theta$ term causes a notorious sign problem\footnote{
In Ref.~\cite{Matsumoto:2021zjf}, we attempted to perform simulations at $\theta=\pi$ using the complex Langevin method with an open boundary condition, which is necessary for the method to work.
However, we found that the topological charge leaks out from the boundary by smearing, which was needed to define the topological charge on the lattice properly.
}, we investigate the phase structure based on analytic continuation from the imaginary $\theta$ region.
This technique has been used for studying the $\theta$ dependence of SU($N$) gauge theory for $N\ge 3$ in the small $\theta$ region \cite{Guo:2015tla, Aoki:2008gv, Panagopoulos:2011rb, DElia:2012pvq, DElia:2013uaf, Bonati:2015sqt, Bonati:2016tvi, Bonanno:2023hhp, Bonanno:2024ggk}.
Similarly, we first calculate the expectation value of the topological charge at imaginary $\theta$ and fit the results to an appropriate holomorphic function. Then $\langle Q\rangle_\theta$ at real $\theta$ is obtained through analytic continuation of the fitting function.
A non-vanishing $\langle Q\rangle_{\theta}$ at $\theta=\pi$ signals spontaneous breaking of CP symmetry.
Unlike the previous studies that used imaginary $\theta$, we employ the stout smearing \cite{Morningstar:2003gk} in defining the $\theta$ term in the action to be used in our simulation.
This is crucial for our purpose because the CP symmetry at $\theta=\pi$ assumes that the topological charge takes integer values, which is not the case if one does not use smearing techniques dynamically.

We find that the spontaneous breaking of CP symmetry occurs at low temperature.
We also find that, as we increase the temperature, the order parameter $\langle Q\rangle_{\theta=\pi}$ decreases and vanishes at some temperature $T_{\mathrm{CP}}$ close to $T_{\mathrm{dec}}(0)$.
We also estimate the deconfining temperature $T_{\mathrm{dec}}(\theta)$ at real $\theta$ by analytic continuation, and find that $T_{\mathrm{dec}}(\pi)  < T_{\mathrm{dec}} (0)$, which is consistent with the general expectation \cite{Unsal:2012zj, Poppitz:2012nz, Anber:2013sga, DElia:2012pvq, DElia:2013uaf, Otake:2022bcq, Borsanyi:2022fub}. 
Thus, our results suggest the relation
\begin{equation}
   T_{\mathrm{CP}} > T_{\mathrm{dec}}(\pi) \ ,
\end{equation}
unlike in the large-$N$ case.

\section{\texorpdfstring{$\theta$}{theta} term in SU(\texorpdfstring{$N$}{N}) Yang-Mills theories}
4D SU(2) gauge theory with the $\theta$ term is defined by the partition function
\begin{equation}
    Z_\theta=\int \mathcal{D}A_\mu e^{-S_{\rm g} - i\theta Q},
    \label{partition-fn-theta}
\end{equation}
where $S_{\rm g}$ is the action for the gauge field $A_\mu$, and $Q$ is the topological charge which is defined by
\begin{equation}
   Q=\frac{1}{32\pi^2}\int d^4x \epsilon_{\mu\nu\rho\sigma} {\rm Tr}\left[ F_{\mu\nu} F_{\rho\sigma} \right].
\end{equation}
The topological charge takes integer values on a 4D torus, which implies that the theory has $2\pi$ periodicity under $\theta \to \theta + 2\pi$.
As a consequence of this fact, the theory has the CP symmetry not only at $\theta = 0$ but also at $\theta=\pi$.

Since the topological charge $Q$ is a CP odd quantity, it is discontinuous at $\theta=\pi$ when the CP symmetry is spontaneously broken.
Therefore, the order parameter of the SSB of the CP symmetry at $\theta=\pi$ is defined by
\begin{equation}
  O = \lim_{\epsilon\to0} \lim_{V \to \infty} \frac{\Braket{Q}_{\theta=\pi-\epsilon}}{V} \,
\end{equation}
where $V$ is the space-time volume. 
Since the Boltzmann weight has a phase factor $e^{-i\theta Q}$, there is a sign problem, which prevents us to measure the order parameter directly.
Instead, we estimate the order parameter through imaginary $\theta$ simulations. 
Note that in the imaginary $\theta$ cases, the phase factor becomes a weight $e^{\tilde{\theta}Q}$ where $\tilde{\theta}=-i\theta \in \mathbb{R}$, hence no sign problem.

There are two simplified cases in which we can obtain the topological charge as a function of $\theta$ explicitly.
One is the dilute instanton gas approximation (DIGA) \cite{Gross:1980br,Weiss:1980rj},
which is expected to be valid at sufficiently high temperature.
The other is the large-$N$ limit of SU($N$) Yang-Mills theory \cite{Witten:1980sp,Witten:1998uka} at low temperature.
The quantity $\Braket{Q}_\theta$ in these two cases is given as
\begin{equation}
    \frac{\Braket{Q}_\theta}{V} = 
    \left\{
        \begin{array}{ll}
          i  \chi_0  \sin \theta &:\ \mbox{DIGA (high $T$)}\   ,\\
        i \chi_0  \, \theta &:\ \mbox{large-$N$ limit (low $T$)}  \ ,
        \end{array}
    \right.
\end{equation}
for $|\theta|<\pi$, where $\chi_0$ is the topological susceptibility at $\theta = 0$ defined by
\begin{align}
  \chi_0 &=  \left. \frac{i}{V} \frac{\del}{\del \theta}
  \langle Q \rangle _{\theta} \right|_{\theta=0} \ .
  \label{top-susceptibility}
\end{align}
This implies that the CP broken phase appears at low $T$, whereas the CP restored phase appears at high $T$ in these simplified cases.
The expectation value $\Braket{Q}_{i\tilde{\theta}}$ is also obtained in these cases as
\begin{equation}
    \frac{\Braket{Q}_{i\tilde{\theta}}}{V} = 
    \left\{
        \begin{array}{ll}
          \chi_0  \sinh \tilde{\theta} &:\ \mbox{DIGA (high $T$)} \ ,\\
          \chi_0  \, \tilde{\theta} &:\ \mbox{large-$N$ limit (low $T$)} \ .
        \end{array}
    \right.
\label{eq:q_im_theta}
\end{equation}
Therefore, we expect that $\Braket{Q}_{i\tilde{\theta}}/V$ grows exponentially in the CP restored phase, whereas it grows with a power law in the CP broken phase.
This is the basic strategy we adopt in this work.

\section{Lattice regularization and the setup for simulations}
In this work, we use the Wilson action defined by
\begin{equation}
S_{\rm g}[U]=-\frac{\beta}{2N}\sum_{n}\sum_{\mu\neq\nu}\mathrm{Tr}\left[ P_{n}^{\mu\nu} (U) \right],
\end{equation}
where $\beta = 2/g_0^2$ and the plaquette is defined by
\begin{equation}
P_{n}^{\mu\nu}(U)=U_{n,\mu}U_{n+\mu,\nu}U_{n+\nu,\mu}^{\dagger}U_{n,\nu}^{\dagger}.
\end{equation}
As the topological charge, we adopt the clover-leaf definition \cite{DiVecchia:1981aev}
\begin{equation}
Q[U]=-\frac{1}{32\pi^{2}}\sum_{n}\frac{1}{2^{4}}\sum_{\mu,\nu,\rho,\sigma=\pm1}^{\pm4}\tilde{\epsilon}_{\mu\nu\rho\sigma}\mathrm{Tr}\left[P_{n}^{\mu\nu}P_{n}^{\rho\sigma}\right],
\end{equation}
where $\tilde{\epsilon}_{\mu\nu\rho\sigma}$ are an anti-symmetric tensor that satisfies $\tilde{\epsilon}_{(-\mu)\nu\rho\sigma}=-\epsilon_{\mu\nu\rho\sigma}.$

\begin{figure}[tb]
\centering
\includegraphics[scale=0.35]{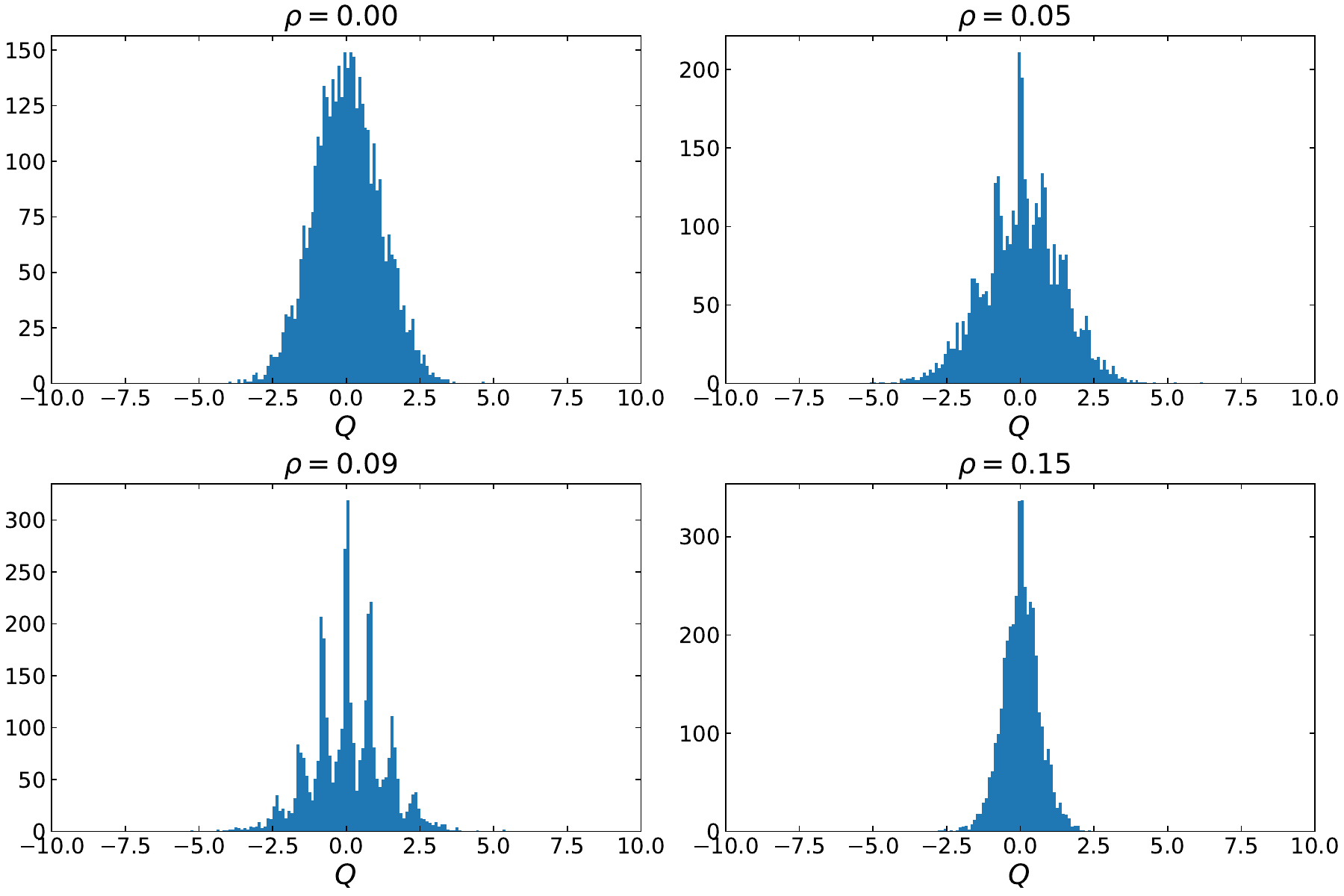}
\caption{The histogram of the topological charge after the stout smearing
for $\rho=0$, 0.05, 0.09, and 0.15 with the fixed number of
smearing steps $N_{\rho}=40$.
The lattice volume is $V_{\rm L}=20^3\times 5$, and
the temperature is $T=1.2 \, T_{\mathrm{c}}$, where $T_{\rm c}$ is the deconfining temperature at $\theta=0$ in the continuum limit.}
\label{fig:Q_hist}
\end{figure}

In order to remove the UV fluctuations from the gauge configuration, we use the stout smearing method \cite{Morningstar:2003gk}.
For the definition of the topological charge which appears in the $\theta$ term, we use the smeared link which is denoted as $\tilde{U}$.
Therefore, for the lattice simulations, we use the following action
\begin{equation}
    S = S_{\rm g}[U] + \tilde{\theta}~Q[\tilde{U}].
\label{lattice_action}
\end{equation}
In our simulations, we set the smearing step size to 0.09 and the number of smearing steps to 40, because this choice enables us to see a comb-like structure in the topological charge distribution. (See Figure \ref{fig:Q_hist}.) As one can see from this figure, the peak positions of the topological charge distribution are shifted due to lattice artifacts. It is well known that one can make the peak positions closer to integers by rescaling the topological charge $Q \to wQ$ with some parameter $w$. (See, for instance, Refs.~\cite{DelDebbio:2002xa, Bonati:2015sqt}.)
Here we determine $w$ so that the cost function
\begin{equation}
  F(w)=\Braket{1-\cos(2\pi wQ[\tilde{U}])}
\end{equation}
is minimized, where $\Braket{\cdots}$
represents the ensemble average.
In our simulations, the topological charge
is defined by $\tilde{Q} \equiv w Q[\tilde{U}]$
with the smeared link $\tilde{U}$.
We will denote $\tilde{Q}$ simply as $Q$ in what follows.

As an algorithm for updating the gauge configuration, we adopt the Hybrid Monte Carlo (HMC) method.
The drift force is coming not only from the gauge part but also from the $\theta$ term when $\tilde{\theta}\ne0$.
Since we use the smeared link for the topological charge in eq.(\ref{lattice_action}), we have to calculate the drift force for the original link by reversing the smearing steps.
For the details, see appendix A of Ref.~\cite{Hirasawa:2024fjt}.

\section{Results}
In this section, we present our numerical results.
We perform the simulations with $L_{\rm s} = 16, 20, 24$ to take the infinite volume limit while $L_t=5$ is kept fixed. We change the temperature by changing the gauge coupling constant using the relationship given in Ref.~\cite{Engels:1994xj}.

To study the $\theta$ dependence of the topological charge, we measure $\Braket{Q}_{i\tilde{\theta}}$ at various values of $\tilde{\theta}$, and then we perform the infinite volume extrapolation and fit the extrapolated value to some holomorphic functions of $\tilde{\theta}$. We estimate the real $\theta$ dependence of the topological charge by performing analytic continuation of the fitting function.
To study the $\theta$ dependence of the deconfining temperature, we measure the Polyakov loop susceptibility in the imaginary $\theta$ region and perform analytic continuation as well.

\subsection{\texorpdfstring{$\theta$}{theta} dependence of the topological charge \texorpdfstring{$\Braket{Q}_\theta$}{<Q>theta}}
\begin{figure}[tb]
    \centering
    \includegraphics[width=0.45\hsize]{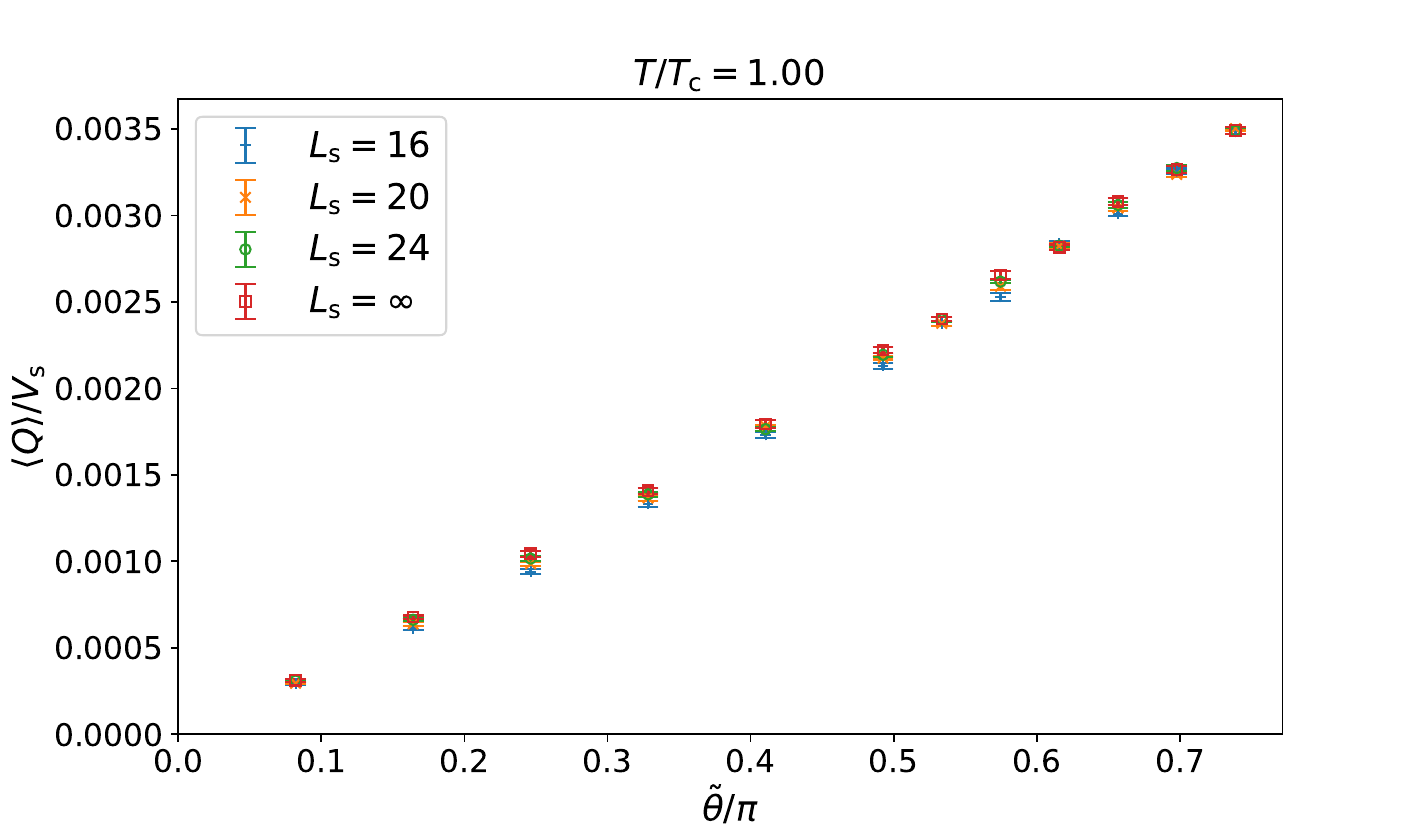}
    \includegraphics[width=0.45\hsize]{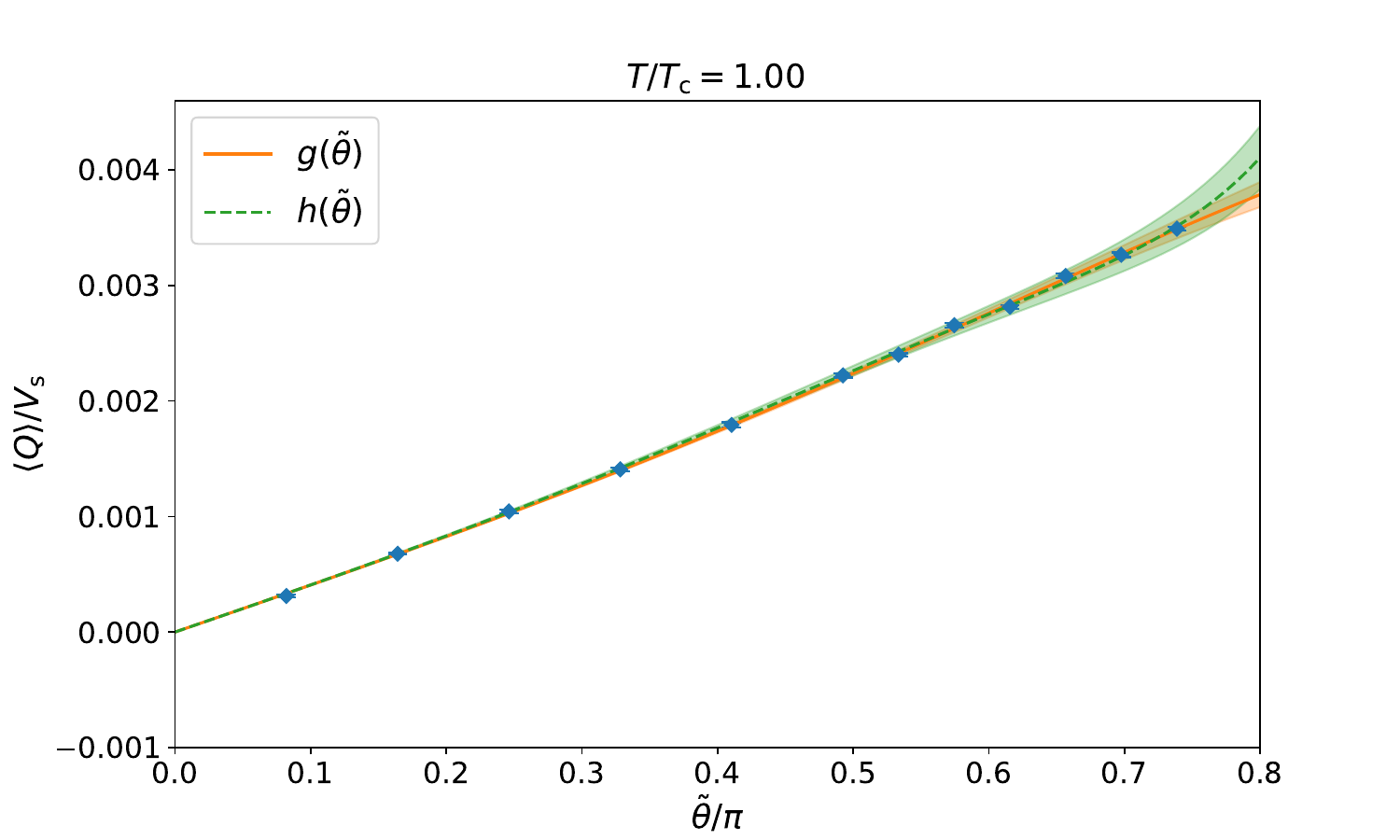}\\
    
    \caption{(Left) $\Braket{Q}_\theta / V_{\rm s}$ is plotted against $\tilde{\theta}/\pi$ for $V_{\rm s}= 16^3,\ 20^3,\ 24^3\ {\rm and}\ \infty$ at $T=T_{\rm c}$. (Right) $\Braket{Q}_\theta / V_{\rm s}$ after the infinite volume limit is fitted to the forms (\ref{g-fitting}) and (\ref{h-fitting}).}
    \label{fig:fitting}
\end{figure}

\begin{figure}[tb]
    \centering
    \includegraphics[width=0.5\hsize]{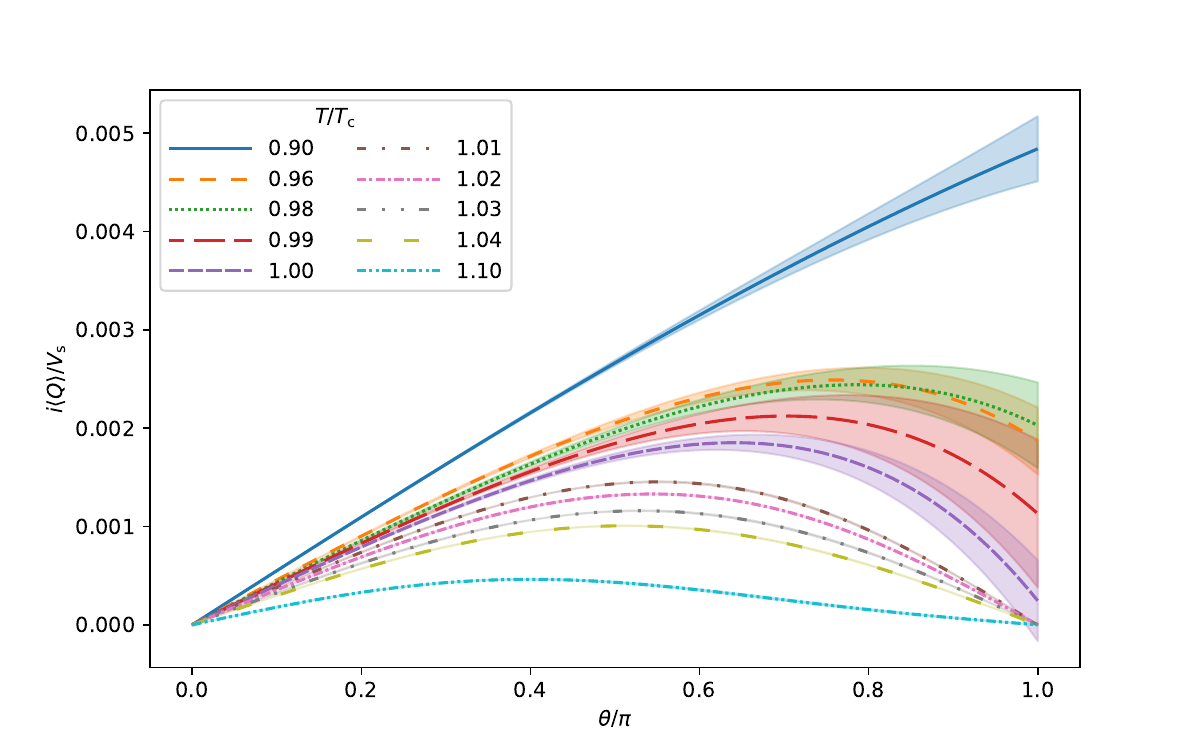}
    \caption{Our prediction for
      $\lim_{V_{\rm s} \rightarrow \infty} i \Braket{{Q}}_{\theta}/V_{\rm s}$
      is plotted against $\theta/\pi$ at various temperature
      within $0.9 \le T/T_{\mathrm{c}} \le 1.1$.
      The gap at $\theta = \pi$ disappears
      at some $T$
      within $1.0 \lesssim T/T_{\mathrm{c}} \lesssim 1.01$.}
    \label{fig:ana_con_poly_det_Tcp}
\end{figure}

First, we take the infinite volume limit of the topological charge density $\Braket{Q}_\theta / V_{\rm s}$ with $V_{\rm s} = L_{\rm s}^3$ at each temperature and at each $\tilde{\theta}$ using the fitting function $f(V_{\rm s})=a_0 + a_1 V_{\rm s}^{-1}$. In Figure \ref{fig:fitting} (Left), we plot $\Braket{Q}_\theta / V_{\rm s}$ for $V_{\rm s}= 16^3,\ 20^3,\ 24^3\ {\rm and}\ \infty$ at $T=T_{\rm c}$ as an example, where $T_{\rm c}$ is the deconfining temperature at $\theta=0$ in the continuum limit. We find that there is no strong finite volume effect at each temperature in $0.9 \le T/T_{\rm c} \le 1.1$.
Then we fit the data to the following forms for the imaginary $\theta = i\tilde{\theta}$
\begin{align}
  g(i \tilde{\theta}) &= \chi_0 \tilde{\theta}
  - a_3 \tilde{\theta}^3 + a_5 \tilde{\theta}^5 \ ,
\label{g-fitting}
  \\
    h(i \tilde{\theta})   &= (\chi_0 - 2b_2 - 3b_3) \sinh{\tilde{\theta}} +
    b_2 \sinh{2\tilde{\theta}} + b_3 \sinh{3\tilde{\theta}}  \ ,
    \label{h-fitting}
\end{align}
which generalize the known results (\ref{eq:q_im_theta}). Note that the coefficients of the first terms are fixed by the relation (\ref{top-susceptibility}).
We also estimate $\chi_0$ in the infinite volume limit at various temperature in $0.9 \le T/T_{\rm c} \le 1.1$, and they are presented in Table \ref{tab:chi0}.
In Figure \ref{fig:fitting} (Right), we plot the fitting curves for $\Braket{Q}_\theta / V_{\rm s}$ after the infinite volume limit against f$\tilde{\theta}$ at $T=T_{\rm c}$ as an example.
We perform the same analysis at each temperature and determine the fitting parameters. 
After the fitting, we perform the analytic continuation of the fitting functions and plot them in Figure \ref{fig:ana_con_poly_det_Tcp}. Here we plot $g(\theta)$ for $T \le T_c$ and $h(\theta)$ for $T > T_c$ since $g(\pi) < 0$ is not consistent with DIGA at higher temperature, which is presumably due to the truncation of the polynomial expansion.
Based on this analysis, we conclude that $T_{\rm CP} \sim T_{\rm c}$.

\begin{table}[t]
  \centering
 \begin{minipage}[b]{0.48\columnwidth} 
    \centering
    \begin{tabular}{|c||c|c|}\hline
        $T/T_{\mathrm{c}}$ & $\chi_0$ & $\chi^2/ N_{\rm DF}$  \tabularnewline\hline 
        \hline 
        0.90 & 0.000351(1) & 0.60 \tabularnewline\hline 
        0.96 & 0.000291(1) & 2.37 \tabularnewline\hline 
        0.98 & 0.000277(2) & 0.53 \tabularnewline\hline 
        0.99 & 0.000268(1) & 0.75 \tabularnewline\hline 
        1.00 & 0.000258(2) & 1.67 \tabularnewline\hline 
    \end{tabular}
 \end{minipage}
 \hspace{-0.1\columnwidth}
 \begin{minipage}[b]{0.48\columnwidth}
  \centering
    \begin{tabular}{|c||c|c|}\hline
        $T/T_{\mathrm{c}}$ & $\chi_0$ & $\chi^2/ N_{\rm DF}$  \tabularnewline\hline 
        \hline 
        1.01 & 0.000243(1) & 2.12 \tabularnewline\hline 
        1.02 & 0.000228(2) & 2.25 \tabularnewline\hline 
        1.03 & 0.000209(2) & 2.91 \tabularnewline\hline 
        1.04 & 0.0001917(9) & 1.78 \tabularnewline\hline 
        1.10 & 0.0001171(5) & 0.98 \tabularnewline\hline 
    \end{tabular}
\end{minipage}
    \caption{The topological susceptibility $\chi_0$ in the lattice unit
      after the infinite volume extrapolation for various
      temperature within $0.9 \le T/T_{\mathrm{c}} \le 1.1$.}
    \label{tab:chi0}
\end{table}

\subsection{\texorpdfstring{$\theta$}{theta} dependence of the deconfining temperature \texorpdfstring{$T_{\rm dec}(\theta)$}{Tdec(theta)}}
\begin{figure}[tb]
    \centering
    \includegraphics[width=0.55\hsize]{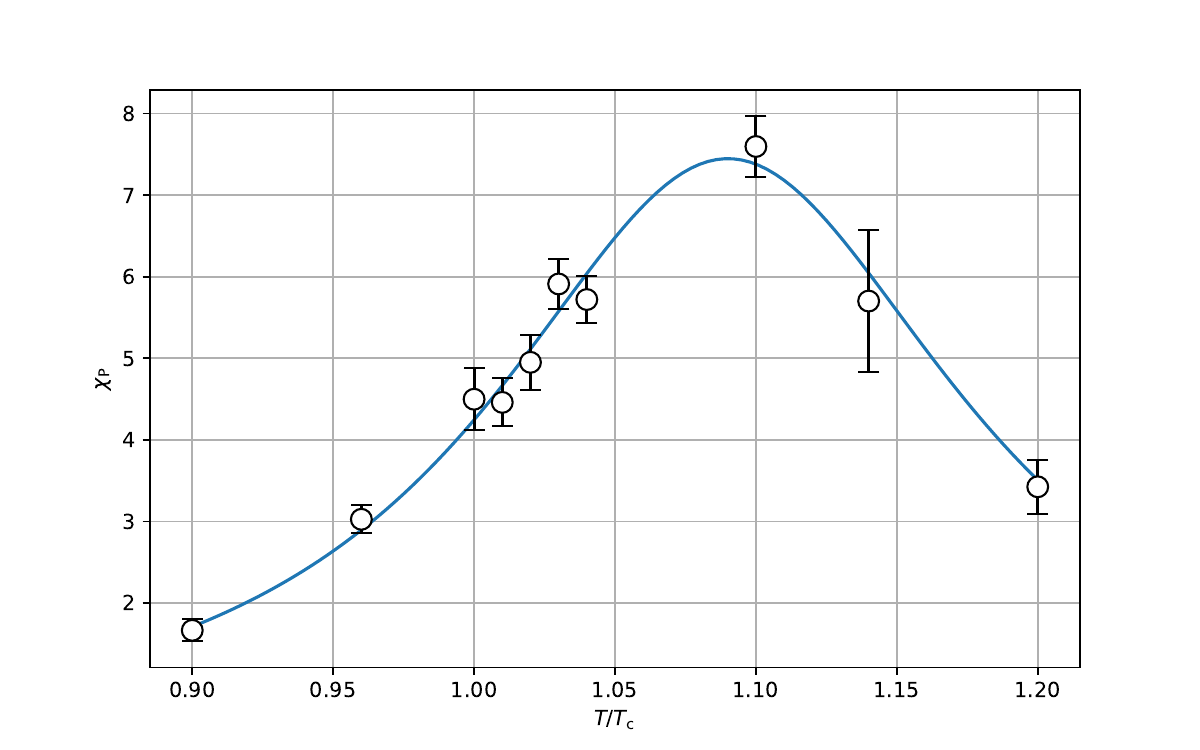}
    \caption{Susceptibility of Polyakov loop $\chi_{\rm P}$ is plotted against $T/T_{\rm c}$ at $\tilde{\theta} = \pi / 2$ for $L_{\rm s}=16, L_{\rm t}=5$. The curve represents the fit to the Lorentz function (\ref{eq:Lorentz_func}).}
    \label{fig:chi_P}
\end{figure}

In order to determine the critical temperature of the deconfing phase transition, we measure the Polyakov loop susceptibility defined by $\chi_{\rm P}(T) = \braket{P(T)^2} - \braket{P(T)}^2$, where $P(T)$ is the Polyakov loop. 
Since the transition is of the second order, $\chi_{\rm P}(T)$ has a peak at some temperature that corresponds to the critical point.
In Figure \ref{fig:chi_P}, we plot $\chi_{\rm P}(T)$ against $T$ at $\tilde{\theta} = \pi / 2$ for $L_{\rm s}=16, L_{\rm t}=5$ as an example.
To determine the critical temperature, we fit the data points to the Lorentzian function
\begin{equation}
    \chi_{\rm P}(T) = \frac{A}{(T-T_{\mathrm{peak}})^2+w^2} \ ,
\label{eq:Lorentz_func}
\end{equation}
where $A$, $T_{\mathrm{peak}}$ and $w$ are the fitting parameters. By extrapolating $T_{\mathrm{peak}}$ obtained at each $L_{\rm s}$ to $V_{\rm s} \to \infty$ using a linear function of $V_{\rm s}^{-1}$, we get the critical temperature $T_{\rm dec}(i\tilde{\theta})$ in the infinite volume limit.

\begin{figure}[tb]
    \centering
    \includegraphics[width=0.55\hsize]{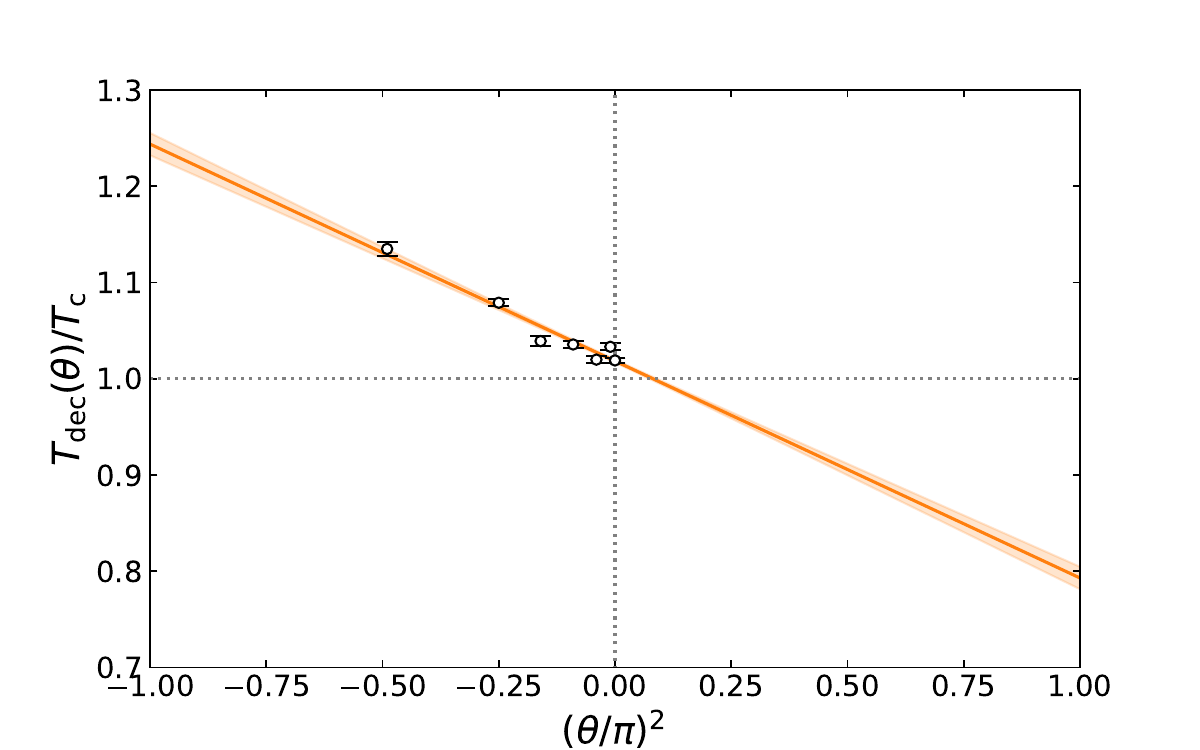}
    \caption{The deconfining temperature after the infinite volume extrapolation
      is plotted against $(\theta/\pi)^2$.
      The line with error band represents the fit to the function \eqref{eq:fit_Tdec}.}
    \label{fig:T_dec}
\end{figure}

In Figure \ref{fig:T_dec}, the results of $T_{\rm dec}(i\tilde{\theta})/T_{\rm c}$ is plotted against $(\theta/\pi)^2$. We fit the data points to 
\begin{equation}
  \frac{T_{\mathrm{dec}}(\theta)}{T_{\mathrm{c}}}
  = c_0 - c_2 \left(\frac{\theta}{\pi}\right)^2 \ ,
\label{eq:fit_Tdec}
\end{equation}
which yields $c_0=1.0183(16)$, $c_2=0.225(12)$\footnote{
In \cite{Yamada:2024pjy}, the deconfinement temperature is obtained at real $\theta$ using the subvolume method. The results lie close to the fitting curve in our figure~\ref{fig:T_dec}.
}.
By extending the fitting function to the real $\theta$ region, we find that the deconfining temperature at $\theta = \pi$ is lower than $T_{\rm c}$.

\section{Summary}
We have investigated the phase structure of pure SU(2) Yang-Mills theory at $\theta=\pi$ based on analytic continuation from the imaginary $\theta$ region.
In particular, we have focused on the relation between the deconfining temperature and the CP restoration temperature at $\theta=\pi$.
The crucial point for our purpose is that stout smearing is performed dynamically.
As a result, the topological charge in the action takes near integer values, which is important for the model to have CP symmetry at $\theta=\pi$.

To determine the CP restoration temperature $T_{\rm CP}$, we have measured the imaginary $\theta$ dependence of the topological charge by fitting the results to the forms (\ref{g-fitting}, \ref{h-fitting}) which generalize the results from simplified models.
Then, by performing analytic continuation of the fitting functions, we find that the topological charge $\Braket{Q}_{\theta}$ at $\theta=\pi$ vanishes at some point in temperature close to the deconfinig temperature at $\theta=0$. 
This result suggests that $T_{\rm CP} \sim T_{\rm dec}(0)$.

On the other hand, we determine the deconfining temperature by measuring the Polyakov loop susceptibility, which has a peak at the critical temperature since the deconfining transition is of the second order in SU(2) Yang-Mills theory.
By performing the analytic continuation, we have found that the deconfining temperature is a decreasing function of real $\theta$. 
Therefore, our result suggests $T_{\rm dec}(0) > T_{\rm dec}(\pi)$, which is consistent with the general expectation \cite{Unsal:2012zj, Poppitz:2012nz, Anber:2013sga, DElia:2012pvq, DElia:2013uaf, Otake:2022bcq, Borsanyi:2022fub}.

By combining these two observations, we can conclude that $T_{\rm CP} > T_{\rm dec}(\pi)$, which suggests the existence of a CP-broken deconfined phase -- unlike the situation at large $N$.
We are also trying to apply this method to 4D SU(3) YM theory to see a possible qualitative difference between $N= 2$ and $N=3$, which has been seen in the case of supersymmetric SU($N$) Yang-Mills theory \cite{Chen:2020syd}.

\acknowledgments
We would like to thank Kohta Hatakeyama for his participation at the earlier stage
of this work.
The authors are also grateful to
Yuta Ito and Yuya Tanizaki for valuable discussions and comments.
The computations were carried out on Yukawa-21 at YITP in Kyoto University 
and the PC clusters at KEK Computing Research Center and KEK Theory Center.
This work also used computational resources of supercomputer NEC
SX-Aurora TSUBASA provided by the Particle, Nuclear, and Astro Physics
Simulation Program No.2020-009 (FY2020), No.2021-005 (FY2021),
No.2022-004 (FY2022), and 2023-002(FY2023) of Institute of Particle and
Nuclear Studies, High Energy Accelerator Research Organization (KEK).
M.~Honda is supported by JST PRESTO Grant Number JPMJPR2117, JST CREST
Grant Number JPMJCR24I3, JSPS Grant-in-Aid for Transformative Research
Areas (A) ``Extreme Universe" JP21H05190 [D01] and JSPS KAKENHI Grant
Number 22H01222.
A.~M. is supported by JSPS Grant-in-Aid for Transformative Research Areas (A) JP21H05190.
A.~Y. is supported by JSPS Grant-in-Aid for Transformative Research Areas (A) JP21H05191.

\bibliographystyle{JHEP}
\bibliography{ref}

\end{document}